# MoveSteg: A Method of Network Steganography Detection

Krzysztof Szczypiorski and Tomasz Tyl

*Abstract*—This article presents a new method for detecting a source point of time based network steganography - MoveSteg. A steganography carrier could be an example of multimedia stream made with packets. These packets are then delayed intentionally to send hidden information using time based steganography methods. The presented analysis describes a method that allows finding the source of steganography stream in network that is under our management.

*Keywords*—network steganography, information, hiding network, detection of new attacks

## I. INTRODUCTION

STEGANOLOGY is a science which has noted a considerable increase in interest in recent years. It is divided into two areas: steganography and stegoanalysis. The former deals with techniques of hiding information in a sent medium which may be, for instance, a stream of packets. Unlike cryptography, steganography consists in hiding the fact that information is being exchanged. The purpose of the other area mentioned above is to analyze media of various types from the perspective of information hidden in them.

The most popular network steganographic techniques are those using sporadically used IP and TCP header fields [4] or even dedicated payload [2]. A variety of network steganography which is more rarely encountered, but is nonetheless equally interesting, is the one using time dependencies between consecutive packets in a given stream. This class of methods is often based on intentional delaying of some messages in order to modulate intervals between consecutive messages. Delays modulated in this manner may be received and interpreted as a series of bits on the far end of communication link [1]. An example of a method which includes the elements of both techniques mentioned above is a hybrid method referred to as LACK (Lost Audio Packets Steganography) proposed in [3]. It uses time dependencies between packets in a given stream, by intentionally delaying packets to such an extent that at the moment of sending them, they are no longer useful and will be rejected on the transmission receiver's side. This method uses a payload of delayed packets as a steganographic carrier. An example of this may be a multimedia stream using the RTP protocol in which consecutive datagrams are sent every 20 ms, and the intentional delay varies from 30 ms to 70 ms [3].

A considerable majority of scientific papers and dissertations on steganology focuses on newer, more effective and efficient methods of detection of hidden information [2] [4] [7]. This refers to both information hidden in multimedia carriers such as images, movies or music, but also in network traffic. This paper focuses on another aspect of steganoanalysis, namely detection of a point from which hidden information is sent. Almost all newest solutions and security measures in cybersecurity treat protection from threats as a passive act, which consists in building a wall, as effective as possible, between the attackers and the sensitive resources we wish to protect. Using the method described in this paper, it is possible to find a point which is a source of steganography transmission in a network managed by us. This allows us to take appropriate measures at the source.

## II. AN IDEA OF MOVESTEG

### A. Description of the method

The method referred to in this paper was initially described in [6]. The main phenomenon occurring in telecommunication networks, which is used in this method, is blurring time dependencies between consecutive packets in a stream along with the number of passes through subsequent transmission devices on a full path length from the transmission source to its destination. This happens irrespective of the transmission medium, therefore in transmission systems which require very accurate synchronization (e.g. SDH), a strong emphasis is put on very accurate synchronization of devices building a network.

An example of this may be an IP network and a multimedia stream sent through it, using an RTP protocol to transmit sound. Due to a limited network efficiency and capacity, packets sent at equal intervals at the source, reach the addressee with various delays. This is a reason for using buffers on the receiver's side, which allows to compensate for the effects of irregular receiving parts of the signal sent. In the event when the network is heavily overloaded with the packets sent, a situation may occur that a part of a signal reaches the receiver too late that is after a consecutive part has reached it.

The phenomenon of blurring time dependencies may be used to find a source of network steganography. As it is demonstrated in the following part of this paper, by examining a delay between consecutive packets, we are able to state that parameters such as delay minimum value, delay maximum value, delay average value and standard deviation vary depending on the distance from the source. In addition, by analyzing a histogram of packets delays, we are able to assign packets belonging to one stream to one of several groups (the number of groups depends on the type of the steganographic method). By analyzing the distribution of delays, we are able to state where the beginning and the end of the steganographic channel are. An example of a histogram of delays between

Krzysztof Szczypiorski is with the Warsaw University of Technolog, Warsaw, Poland (e-mail: ksz@tele.pw.edu.pl).

Tomasz Tyl is with the Warsaw University of Technolog, Warsaw, Poland (e-mail: ttyl@mion.elka.pw.edu.pl).

consecutive packets in a stream, along with an instruction on how to interpret it, is presented in Section 4 of this paper.

*B. Assumptions and limitations*

A success and efficiency of this method depend on several assumptions which have to be met. The first one is that it is necessary to analyze every stream between any pair of communicating hosts, as a separate stream. Information, that is the source host, the target host, the sequence number, and the time of receiving the last packet, must be stored for every stream flowing through every node participating in the measurements. Another assumption which has to be met is that there has to be one central place in which all the paths in our network are known, which allows to correlate results from the entire route of a given stream used as a steganographic carrier. Another assumption, equally important to the previous one, is an ability to detect steganography via nodes used for analysis. The last assumption is a possibility to accurately measure the time interval between consecutive packets of a given stream. Depending on the frequency of sending packets used as a steganographic carrier, measuring time with appropriate accuracy is necessary.

This method has its several limitations which decrease its efficiency or even make it impossible to use it. The first and most important limitation to the method is the type of steganography detected. The method works only for steganography using time dependencies between consecutive stream packets, irrespective of whether these are methods directly using delays or the hybrid ones. Another limitation is a negative impact of multipath traffic on the method efficiency. This means that if a selected steganography method allows for a transmission via several separate paths (for in-stance, a steganographic carrier is several parallel multimedia streams, each of which is sent via different route), such traffic is much more difficult to analyze. An accurate synchronization and correlation of measurements taken from various nodes is then necessary, which is not always possible. The last limitation worth mentioning is the impact of nodes disturbing time dependencies on the method efficiency. As demonstrated in the next section of the paper, nodes which modify delays (for instance, as a result of their overload), cause total blurring of time relations between the packets.

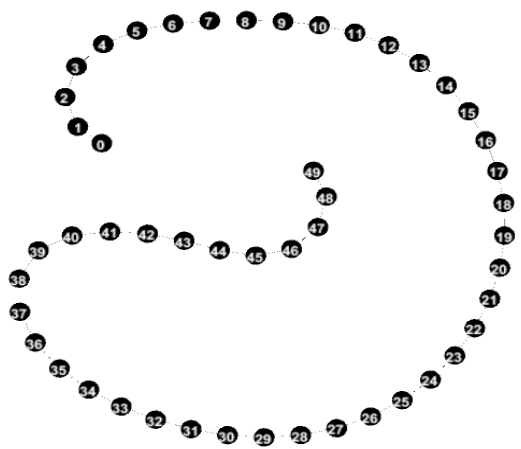

Fig. 1. Line topology consisting of 50 hosts

### III. SIMULATION ENVIRONMENT

In order to examine the efficiency of the method presented in the previous section, a simulation environment was created, allowing to dynamically define input parameters and a topology of the network used. The environment consists of a group of virtual machines which are managed by one central server connected with a database storing measurement results. VirtualBox 5.0 was used as virtualization software, the managing server was written in Python 2.7, the database engine is MySQL Community 5.7, the system of virtual machines is Alpine Linux 3.2. The entire environment was launched on one physical machine with a quad core 3.2 GHz processor and 12 GB of RAM, an operating system launched on this machine was Ubuntu 14.04 LTS.

The main and most important part of the environment was the server which dealt with the setup of hosts and a virtualizer, but was also responsible for defining experiments, sending commands and collecting results of a completed test. The virtual machines were set up to send traffic in line with the routing table known to them, obtained through an OSPF protocol launched on every host. The network setup in the virtualizer involved creating a relevant number of private networks consisting of exactly two interfaces, one from each directly connected host.

This paper focuses on two network topologies, that is a line and a Manhattan-type network. The former will easily allow to present phenomena occurring in the network when sending network steganography using time dependencies between packets, and the latter will present the impact of network topology complexity on the efficiency of the method. Both topologies referred to above are schematically presented in Fig. 1 and 2.

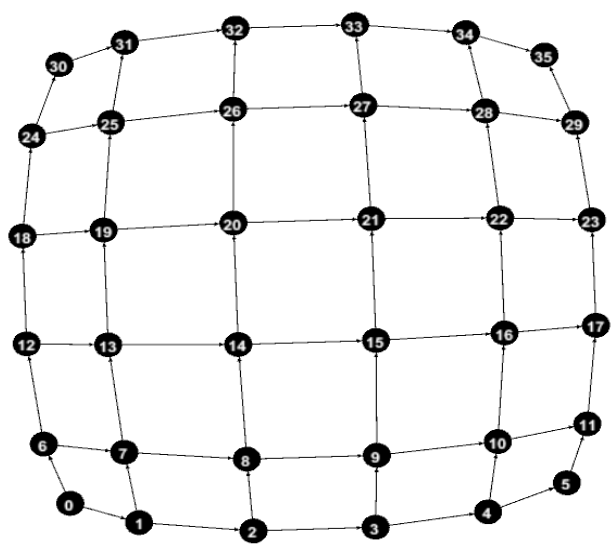

Fig. 2. Manhattan-type topology consisting of 36 hosts

It should be highlighted that for the Manhattan-type topology, the network structure from the perspective of, for instance, host number 0, and taking into account the paths traced by the OSPF protocol, is as depicted in Fig. 3.

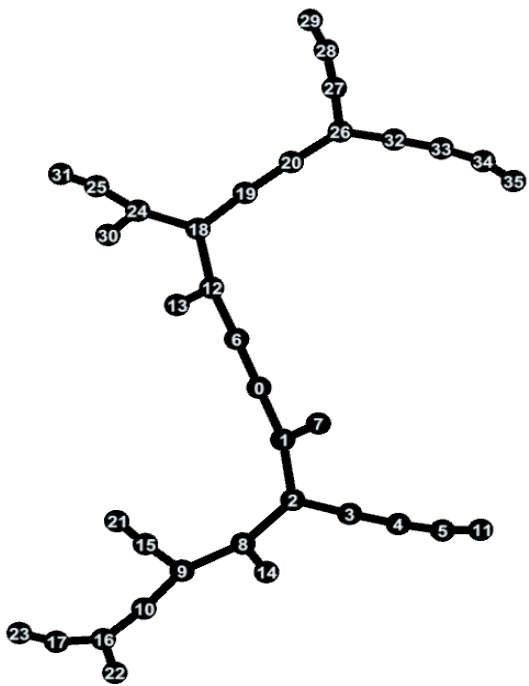

Fig. 3. Manhattan-type topology from the perspective of node 0, taking into account routes traced by a routing protocol, consisting of 36 hosts

Such a transformation of the scheme allows to considerably simplify the traffic analysis.

## IV. RESULTS AND CONCLUSIONS

To depict the options the method provides, simulations of two network topologies described above are presented below, that is a line topology consisting of 50 nodes and a Manhattan-type topology consisting of 36 nodes. The former will allow to explain how the method works in a simple way, and the latter will demonstrate the options and efficiency of the method involving a more complicated network.

### A. Line topology

The first case presented below will use a LACK steganographic method. The source node will be node 0, and the receiver node will be node 49. A steganographic channel is over the entire transmission length. A steganographic carrier is UDP packets sent at the source at equal intervals. All the parameters are presented in Table I.

TABLE I
Input parameters, case no. 1

| Parameter | Value | Unit |
|---|---|---|
| T1 | 20 | ms |
| T2 | 30 | ms |
| P | 15 | % |

The meaning of the parameters is the following: T1 is the nominal value of delay between consecutive packets sent from the source. T2 is a value by which packets used for steganographic transmission are delayed. Both parameters referred to above have been selected according to the values in [3]. The last parameter is P – this is the percentage of traffic carrying hidden information in relation to all packets of a given stream. In a majority of cases, this parameter assumes much lower values within the range of several per cent, but the value assumed will allow to better present how the method works. The simulation duration time for all cases described in this paper was assumed to be 120 seconds.

In the first case, the analysis of time dependencies will be launched in all nodes at the entire route length. Once the simulation has been completed, information on the packets sent is collected in the server which saves it in the database. Next, a histogram of delay between consecutive packets is set for every node, along with the values of minimum, maximum and average delay between consecutive packets; in addition a standard deviation is established. Then, histograms are combined in the relevant order, which allows to see how time dependencies change with the distance increase. The histogram described above is presented below.

The X axis of the histogram contains node numbers, that is values from 1 to 48 (the sending and the receiving node are not included in the traffic analysis). The Y axis contains the value of delay between consecutive packets expressed in microseconds. The Z axis presents the number of packets which match a given range, that is a number of packets whose delay versus the previous packet fell within the range of a given bin. The number of bins in every histogram is 100. The interval of bins for every case described in this paper was established by dividing the difference between the maximum and the minimum value of delay by 100. Groups into which the stream used for steganographic purposes is divided are marked with letters A, B and C. It should be highlighted here that the histogram described above refers to the traffic related to only one UDP stream sent from node 0 to node 49. A group of packets containing hidden information has been market with letter A. These packets are delayed by 10 milliseconds on average with respect to the preceding packet. This is the case because, as it has already been mentioned above, these packets are intentionally delayed by additional 30 milliseconds, so they are sent after their successor already at the source. Sending packets at regular intervals of 20 ms and then delaying one of them by additional 30 ms is a reason for some 10 millisecond delay with respect to the predecessor. In line with the input parameters, in this case this is some 15% of all packets of the stream under analysis. The most numerous group of packets is marked with letter B – these are packets which were not intentionally delayed, therefore their average delay is some 20 ms. In addition, they do not carry stenograms. This group contains some 70% of all packets of the stream. Packets which were sent just before the ones containing hidden information are marked with letter C. Their average delay with respect to their predecessor is some 40 ms, that is twice as much as the value of T1. This group's number corresponds to group A, which is also some 15% of the entire traffic.

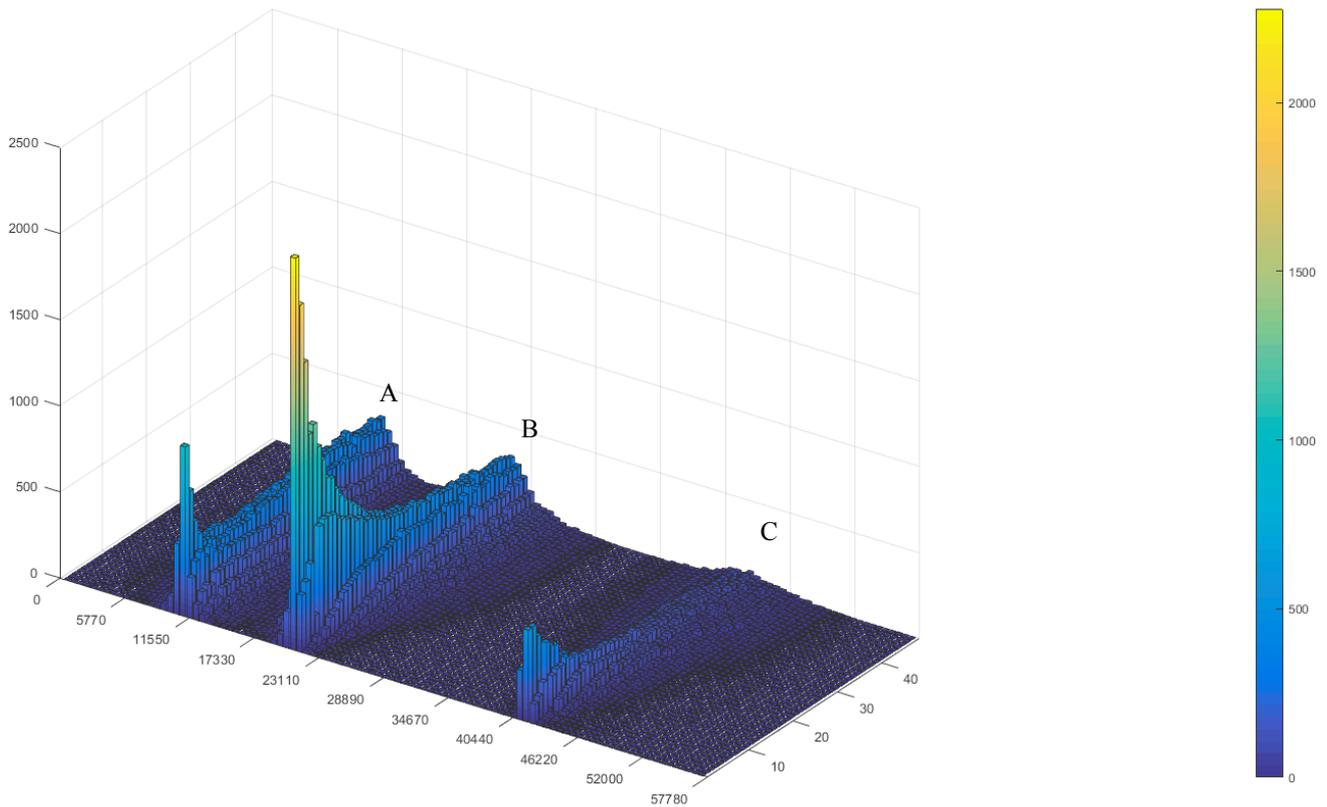

Fig. 4. Histogram of delays between consecutive packets for every node. Case no. 1

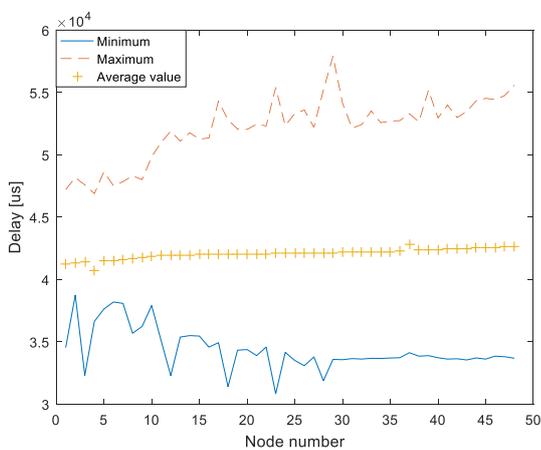

Fig. 5. Plot showing the minimum, maximum and average value in case no. 1

As expected, the minimum value of delay slightly decreases, the maximum delay increases and the average value changes the least and also increases. Movement of the average value towards longer delays is due to the fact that the time of a packet processing in each node, irrespective of the packet, is non-zero. The marginal values (minimum and maximum) go with every consecutive node to the extreme, because when packets are sent at equal intervals at the source, after the first node most of them still preserve time dependence with respect to the neighbouring packets, yet after several subsequent nodes, due to random events occurring in the network and in every node, the dependencies become levelled out. As depicted in the histogram, the distribution of delays resembles a bell in its shape. Initially, the bell is slender and tall, then it becomes lower and broader with the route length. The next plot presents a standard deviation of delay of packets which belong to group A.

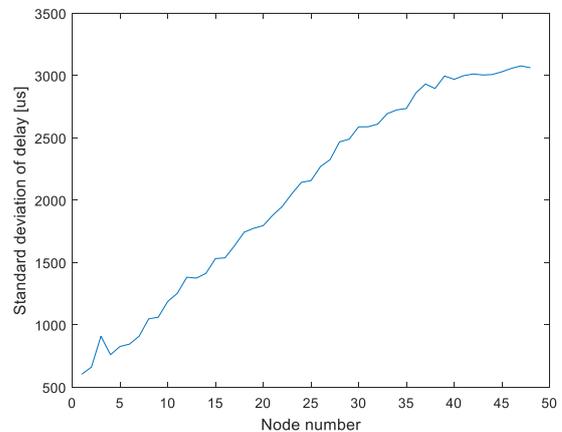

Fig. 6. Standard deviation in case no. 1

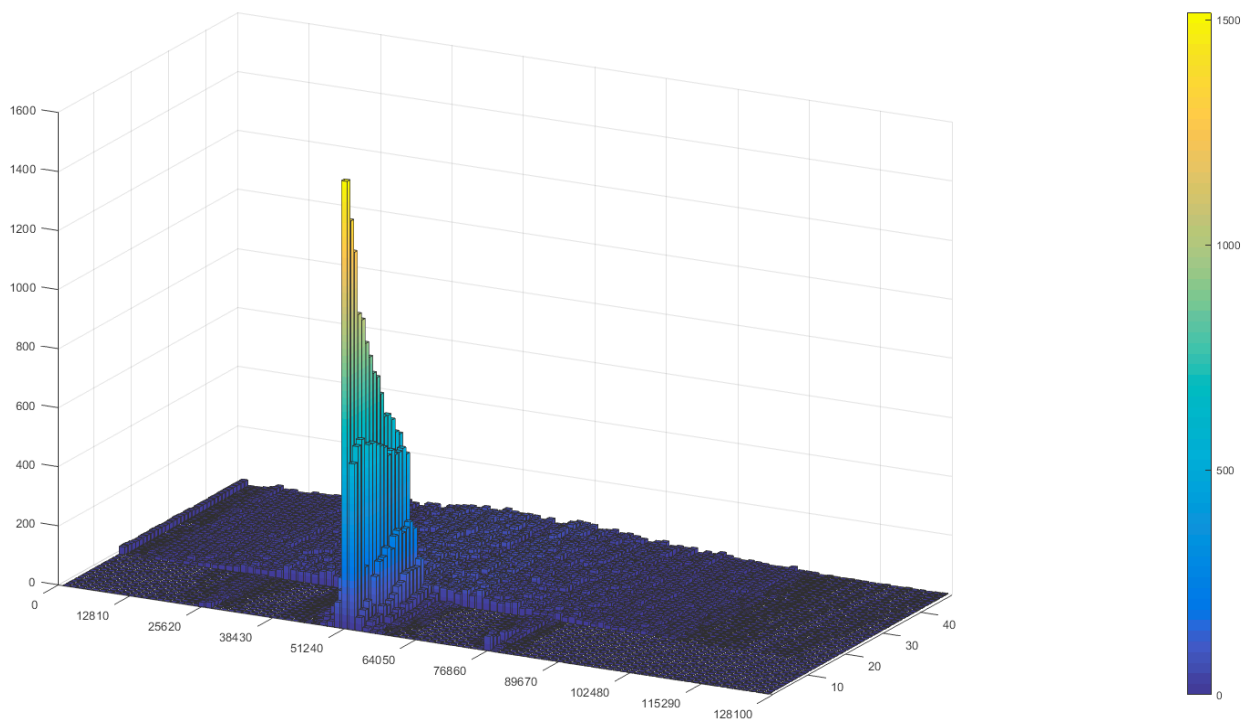

Fig. 7. Histogram of delays between consecutive packets for every node. Case no 2

As presented in the plot above (Fig. 6), the value which changes the most along with the way covered by the packets, is the standard deviation of the delay between consecutive packets. This confirms the change in the shape of the histogram for nodes located at the end of the route.

Another example in this topology is a case using modulation of delay between consecutive packets in order to assign "zeros" and "ones". This method consists in intentional delaying a packet to signal a logical "one" to the party receiving the steganogram, or in intentional advancing (a buffer is required in which we will store a specific number of messages) in order to assign a logical "zero" (assignment of "zeros" and "ones" may be reverse). Similarly to the previous case described above, nodes with numbers from 1 to 48 will be colleting information on time dependencies between consecutive packets. As an additional element in this case, we will introduce unintentional delay in node no. 15. The purpose of this is to simulate an overloaded network node, which inadvertently introduces delays to all packets that pass through it. Delay is added on the outgoing interface of the host with a tool available in Linux, that is Traffic Control [5]. The distribution of the delay is normal, with an average value of 30 ms and a standard deviation of 15 ms. The other input parameters are described in the table II.

Analogically to the previous case, a histogram was calculated for every node, and then the histograms were combined into one three-dimensional histogram presented below in Fig. 7.

TABLE II
Input parameters, case no. 2

| Parameter | Value | Unit |
|---|---|---|
| T1 | 50 | ms |
| T2 | 25 | Ms |
| P | 5 | % |
| L | 100 | |

Similarly to the previous example, the main part of the multimedia stream is initially focused on one value equal to 50 ms, (that is the value of parameter T1). The packets used for modulation of the bits which are a steganogram constitute a much smaller part of the whole. Small peaks are visible near the delays amounting to 25 ms and 75 ms (T1 ± T2). Both peaks comprise 5% of all packets. At node 15 which disturbs time dependencies, we can see that the groups which have been clearly differentiable so far, have been totally mixed with each other, and the number of packets allocated to each bin is similar. The vertical edge visible for delays equal to 0 ms is due to the fact that the Traffic Control tool is unable to accelerate the packets (impose a negative delay on them), all these packets were sent immediately after their arrival in the node.

Let us look now at the diagrams presenting the minimum, maximum and average value of delay between consecutive packets. In this case, all packets of the multimedia stream are taken into consideration.

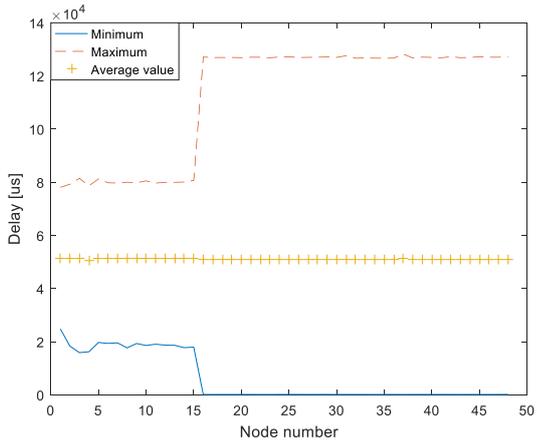

Fig. 8. Diagram presenting the minimum, maximum and average value of case no. 2

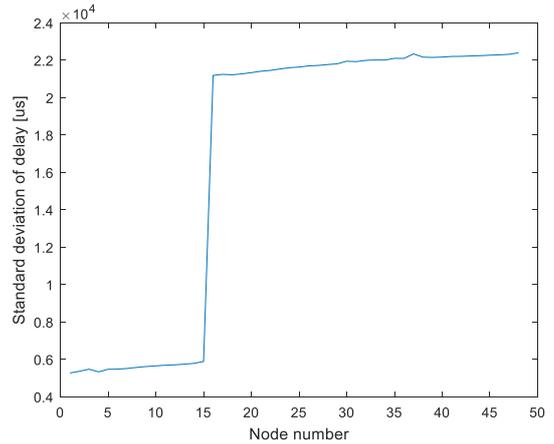

Fig. 9. Diagram of standard deviation in case no. 2

The plot in Fig. 8 presents the extent of impact which strong noises and disturbances created in heavily loaded or damaged network devices have on the method efficiency and a possibility to select the groups which are part of the stream. Below, a diagram of standard deviation is presented, which is also strongly disturbed from node 15 by introduced disturbance of time dependencies.

### B. Manhattan-type topology

A more complicated case, in which network topology was depicted in Fig. 2, will now be presented. The node sending a multimedia (and steganographic) stream is node number 0. The recipients of transmission are two hosts number 5 and 35. Every topology may be presented in a perspective in which a given host can see it, in this case it was depicted in Fig. 3. As it has been stated before, this transformation is a result of paths traced by the OSPF routing protocol. The nodes collecting data on packet delays will be nodes number 1, 6, 7, 3, 9, 15, 21, 20, 19, 18, 34, 28, 29 (the selected nodes create three "walls" between node 0 and 35 in the topology diagram). LACK will be the steganographic method used. An additional element is a disturbing node no. 12 which delays every packet by 30ms ± 15ms with a correlation between the value of the delay added with respect to the previous one equal to 15%. The simulation duration time is identical to the previous ones and amounts to 120 seconds, the other parameters are presented in the table below.

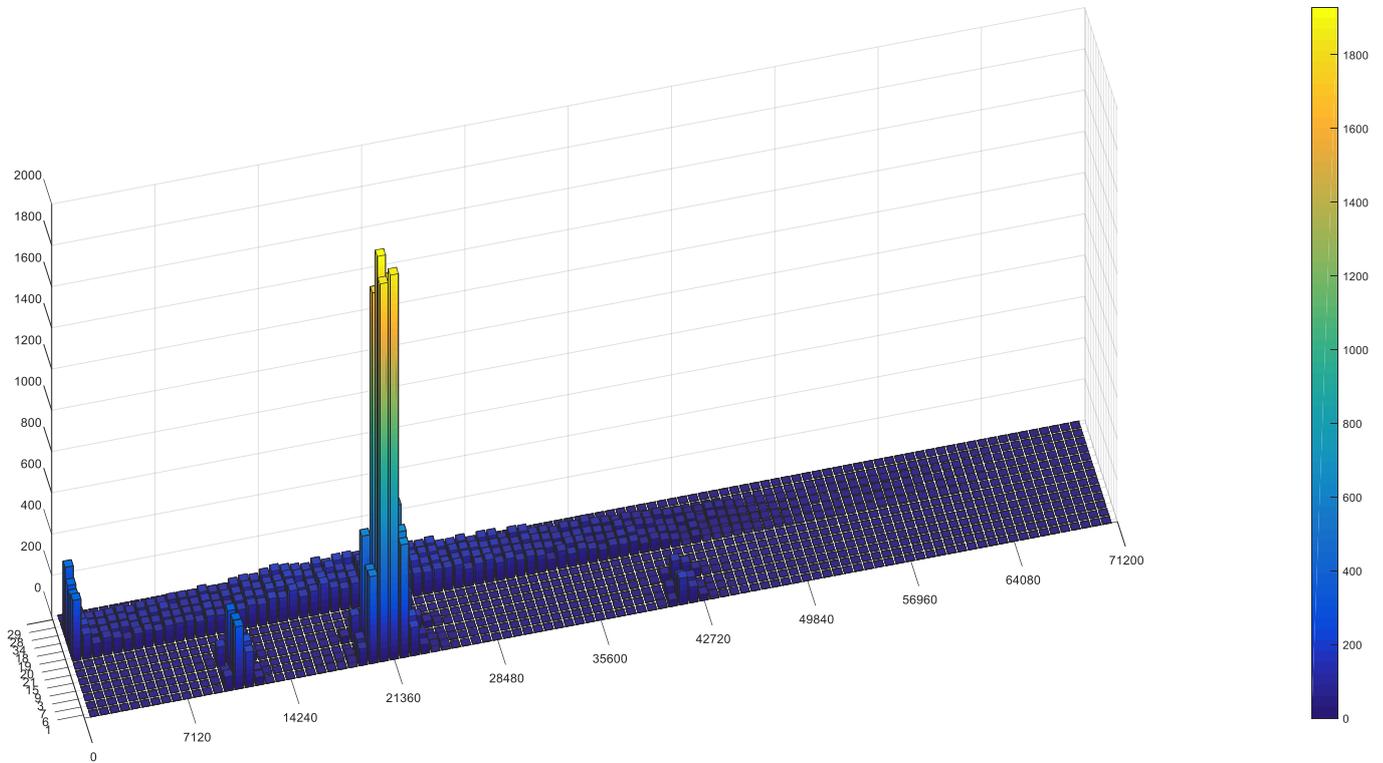

Fig. 10. Histogram of delays between consecutive packets for every node. Case no. 2

TABLE III
Input parameters, case no. 3

| Parameter | Value | Unit |
|---|---|---|
| T1 | 20 | ms |
| T2 | 30 | ms |
| P | 5 | % |

Identically to the two previous cases, a histogram for every network node was created (Fig. 10).

The histogram analysis itself allows to state that in this case it is difficult to unambiguously conclude that time dependencies along with subsequent nodes on the path will be blurring. Several first nodes in which we collect measurements have active probes have a histogram of delays that is very similar, then everything becomes blurred by disturbances introduced in node no. 12. With such a small number of nodes collecting measurements and under the influence of external disturbance, we are unable to conclude where the source of the steganographic stream is, even despite two separate streams, which theoretically should demonstrate that in the proximity of node 0, time dependencies are precisely established, and along with departing from it, both in the direction of node 5 and 35, time dependencies blur (similarly as in the first case presented).

V. CONCLUSIONS

The method MoveSteg previously proposed in [6], in line with what has been described in the previous sections, allows to observe the process of blurring time dependencies between packets. This phenomenon occurs irrespective of whether a given packet is used as a steganographic carrier. The conclusions presented in this paper may serve to detect the source sending steganography in our network.

The scenario of using this method in real world is the following. With a server in our network seized by a criminal (hacker), from which a steganographic stream is sent in several directions via other paths, we are able to find the source of steganography by analyzing the traffic from the sufficiently large number of nodes. The seized server may introduce disturbances to time dependencies not only in streams which originate in it, but also in the ones that pass through it. If we are able to analyze the traffic for every of such passing streams, we will notice groups of packets which were described in Section 4.1. Along with the growing distance from such a server, every group will be undergoing an increasing blurring. By collecting and correlating several such observations, we will be able to indicate which node in the network is conducting steganographic transmission.

A weakness of this method is its dependence on the number of sounds on the path of the stream used as a steganographic carrier. With a small number of hosts collecting measurements, we are unable to state with sufficient certainty, whether along with the increasing distance from the suspicious node, time dependencies in streams become blurred. Another weakness of this method is the analysis sensitivity to disturbances introduced unintentionally by other elements of the network, for instance heavily overloaded or damaged transmission devices. There may be a case when a node suspected to send steganography is in fact an old server with insufficient resources.